# Bolometric response in graphene based superconducting tunnel junctions

Heli Vora, Piranavan Kumaravadivel, Bent Nielsen, and Xu Du[*]

Department of Physics and Astronomy, Stony Brook University

*email: xudu@notes.cc.sunysb.edu

## Abstract

We fabricate graphene-TiOx-Al tunnel junctions and characterize their radio frequency response. Below the superconducting critical temperature of Al and when biased within the superconducting gap, the devices show enhanced dynamic resistance which increases with decreasing temperature. Application of radio frequency radiation affects the dynamic resistance through electronic heating. The relation between the electron temperature rise and the absorbed radiation power is measured, from which the bolometric parameters, including heat conductance, noise equivalent power and responsivity, are characterized.

Nanomaterials such as nanowires and carbon nanotubes have been studied as radiation absorbers for bolometer applications in the recent years, because of their low heat capacitance and therefore fast response[1-5]. More recently the isolation of graphene, a single atomic layer of crystalline carbon and the thinnest solid-state material[6-8], has opened a wide range of opportunities for studying the unique properties of the two-dimensional Dirac fermionic systems[9]. This also opens the possibility of utilizing graphene's unique physical properties for building the state-of-the-art bolometers. Benefited from the chiral Dirac fermionic electronic structures [7-9], the charge carriers in graphene are highly mobile. The density of states (DOS) in graphene depends linearly on Fermi energy and approaches zero at the charge neutral Dirac points where the conduction and valence bands meet. As a result of the small volume and low DOS, graphene has very small electron heat capacitance. Considering the simple case of Dirac electron gas in which $C_e = A \int \varepsilon D(\varepsilon) \frac{df(\varepsilon)}{dT} d\varepsilon$ (here $A$ is the area of graphene, $D(\varepsilon) = \frac{2\varepsilon}{\pi(\hbar v_F)^2}$ is the DOS, and $f$ is the Fermi distribution function), we can estimate the value of the electron heat capacitance in graphene and its gate voltage and temperature dependence, as illustrated in Figure 1b. At experimentally relevant conditions we find the heat capacitance in graphene can easily reach the values which are extremely low (e.g., $C_e \sim 10^{-21}$ J/K for T < 5K at Vg~10V, for a 1μm$^2$ channel) to be achieved in conventional metal structures. Such low heat capacity can be utilized as a significant advantage in bolometer devices whose response time is limited by the ratio of the heat capacitance and thermal conductance: $\tau = C/G$. As proposed for other nanomaterial-based bolometers such as carbon nanotube bolometers[10], having a low heat capacitance allows simultaneous achievements of short response time $\tau$ and low heat conductance $G$, and hence low noise equivalent power (thermal fluctuation limited to $NEP = \sqrt{4k_B T^2 G}$). Another advantage of graphene for bolometer applications, compared with other lower-dimensional nanomaterials such as carbon nanotubes[5], is the large contact area which allows relatively low contact resistance. Since most of these devices will have micrometer sizes, antenna coupling and hence impedance matching are needed for high detection efficiency. Low and gate tunable device impedance in the graphene based bolometer may provide significant advantage over other nano-bolometers [11]. In all, the combination of low heat capacitance and low, gate tunable impedance makes graphene potentially promising in bolometer applications [12], for achieving simultaneously very fast speed (short time constant) and high sensitivity.

At this time very little experimental work has been done on studying the bolometric response of graphene. While in the long term it is desirable to understand the time response and the gate tuning of the graphene bolometers, it is necessary first of all to demonstrate the basic bolometric response of the material under microwave radiation. In this letter we report fabrication and characterization of graphene-aluminum superconducting tunnel junction (STJ) bolometers. The STJ device structure discussed here serves two purposes: 1. to provide a sizable resistance-temperature response. In a typical graphene junction with normal contacts, due to the weak contribution of resistance from electron-phonon scattering, resistance away from Dirac point is rather insensitive to temperature[13], making it difficult to achieve sufficient responsivity. In a graphene STJ, when the junction is at bias $V_b < 2\Delta$ ($\Delta$ being the superconducting gap of the leads), the junction conductance is dominated by the tunneling of the thermally excited quasiparticles, which shows strong temperature dependence; 2. To provide thermal isolation to the electrons within the graphene channel, by significantly cutting down the quasiparticle tunneling from graphene into the superconductor when the Fermi level is within the superconducting gap of the leads. By applying radio frequency (RF) radiation to the devices, we heat up the graphene absorber whose electronic temperature increase results decrease in the tunneling resistance (voltage), which can be measured using a DC setup. By comparing the device characteristics under radiation with those in various bath temperatures absent of radiation, we obtain the temperature increase at corresponding RF power, from which noise equivalent power and responsivity of the graphene STJ bolometers can be calculated[14].

The graphene superconducting tunnel junctions discussed here were fabricated using mechanically exfoliated graphene, on $Si/SiO_2$ (285nm) substrates. The junctions were defined with standard electron beam lithography (EBL). Approximately 2nm titanium was e-beam evaporated, and then briefly oxidized to form semitransparent barriers. Aluminum (~30nm) was then e-beam evaporated as superconducting electrodes, followed by standard lift-off process. Typical devices are 2-4 μm wide and ~0.5μm long. Note that we used semitransparent TiOx barrier instead of the commonly used high quality AlOx barrier mainly because: 1. Ti (and TiOx) has better sticking/wetting property on graphene and therefore is relatively easy to achieve low or zero pinhole density; 2. The large dielectric constant of TiOx (~100) allows large tunnel junction capacitance and hence low impedance for RF signals. This is important both for better impedance matching and for reducing the non-linearity induced response. 3. Semitransparent TiOx barrier allows relatively low junction resistance (typically ~kΩ), which is more convenient to measure in our current setups. However, such non-ideal barrier also leads to undesirably large heat conductance (due to leaking of the hot quasiparticles) and hence high thermal-fluctuation-limited noise equivalent power. Such limitation is not intrinsic to graphene. With better control of the barrier quality and thickness, it is in principle possible to fabricate high quality tunnel barriers to achieve better tunneling characteristic and at the same time reasonably low tunneling resistance.

The devices were measured in a dilution refrigerator. DC transport measurements were performed using a Lock-in amplifier operating at 13Hz, connected to the sample through twisted-pair cables which are filtered with EMI π-filters at room temperature and RC filter (with ~1kHz cut-off frequency) at ~1.5 Kelvin. RF power was delivered using a coaxial cable with open end located about 10cm away from the devices. No antenna was used, because of which only a small fraction of the applied power radiates out and was picked up by the devices. To calibrate the power received by a device, we placed a 50Ω resistor in series in the close vicinity of the graphene STJ device. A RF oscilloscope with 50Ω input impedance was used to measure the voltage across the resistor, from which the current through the sample can be estimated. The modeling of the STJ device was shown in Figure 1a. Each STJ contact is modeled as a resistor (with DC tunneling resistance) in parallel with a capacitor (Al-TO$_X$-Graphene, ~5pF estimated from the geometry of the contacts). The graphene channel is model as a resistor (with DC graphene resistance of $R_{chan}$~200Ω, estimated from the typical resistance of 4-terminal devices with similar aspect ratio) in parallel with a capacitor with source-drain capacitance (note that the source and drain electrodes are coupled capacitively through the conducting backgate, with an estimated capacitance of 2pF). The RF power absorbed for heating the absorber was calculated using the current through the resistive graphene channel as $P=I^2R_{chan}$. It should be noted that the analysis of data involves characterizations using parameters that cannot be precisely measured (but instead estimated by calculation), and is therefore not accurate. However, order-of-magnitude estimations can be drawn in charactering the bolometric response of the STJ devices.

Upon cooling down the samples to below ~1K, we observed characteristics of S-I-N tunnel junctions with non-ideal semitransparent barriers in the dynamic resistance (*dV/dI*) vs. bias voltage dependence[15], as shown in Figure 2a. Here the dynamic resistance was measured by supplying the samples with the combination of a small AC current ($I_{mod}$=10nA) and sweeping offset current, both provided by a Keithley 6221 current source. The AC component of the voltage response was measured using a Lock-In amplifier, from which the dynamic resistance $dV/dI = V_{mod}/I_{mod}$ was calculated. The DC bias was amplified and measured by a DC voltmeter. When the junctions are biased outside the superconducting gap $V_b>2\Delta$, the dynamic resistance is roughly bias voltage independence. For $V_b<2\Delta$, the quasiparticle tunneling is suppressed by the superconducting gap, resulting in increasing of the dynamic resistance. The observed critical bias voltage below which the dynamic resistance increases gives $2\Delta$~0.25mV, consistent with the reported superconducting gap of aluminum. Reducing the temperature induces a significant increase of the dynamic resistance within the superconducting gap, as a result of the reduced thermal excitation and hence quasiparticle tunneling. By measuring the bias voltage dependence of the dynamic resistance at various temperatures in absence of RF radiation, we built up a temperature calibration based on the zero bias dynamic resistance of the STJ.

Next we investigated the heating of electrons through RF power. At the base temperature of 160mK, we applied RF power at a fix frequency of 600MHz. The amplitude of the RF power was limited so that the bath temperature measured by a thermometer showed no observable increase. For absorbed power of P < -90dBm, we observed no significant change in the *dV/dI* vs. $V_b$ dependence. Between P = -90 ~ -55 dBm, a decrease of tunneling resistance within $V_b<2\Delta$ was observed with increasing applied RF power (Figure 2b), similar to that observed with increasing bath temperature. Above P = -50 dBm on sample, we started to observe an increase in bath temperature. Here we limit our discussion to the range of P = -90 ~ -55 dBm.

To understand the nature of the RF radiation induced dynamic resistance change, we compare the *dV/dI* vs. $V_b$ curves measured under RF radiation with those measured in absence of RF radiation but at higher bath temperatures. Figure 2c shows such comparison between two measurements: T=160mK/-70dBm vs. T=500mK/no-RF. The near-perfect overlap of the two curves strongly indicated that RF radiation impact on the devices was predominantly through electronic heating. The slight difference may be attributed to the small difference in the superconducting gap for these 2 situations, resulting from the different bath temperatures and hence the temperature of the superconducting leads. This hypothesis is supported by the comparison of the curves at stronger radiation power/higher temperature, such as shown in Figure 2d (160mK/-60dBm vs. 950mK/ no RF). Here we see a clear difference between the 950mK/no RF curve and the 160mK/-60dBm curve in that the former has a narrower peak, as a result of the significantly reduced superconducting gap at the temperature close to $T_c$. The comparisons here suggest that the RF radiation drives the electrons in graphene to an effective temperature above that of the substrate, as the superconducting leads stay at the bath temperature.

We note that another possible mechanism for "smearing" the dynamic resistance peak inside the superconducting gap is that the RF amplitude effectively averages over some region of the IV curve. The low frequency Lock-in amplifier measurements will be affected by such averaging if the IV curve is non-linear and the RF voltage drop is large (compared to the superconducting gap in this case). The resulting Lock-in-measured dynamic resistance vs. bias dependence can be calculated based on the IV curves taken without RF radiation, as shown in the inset of Figure 2b. It can be clearly seen that the non-linearity induced RF smearing does not agree with our observations. The reason for the negligible non-linearity effect in our measurements is that the RF signal sees effectively low junction impedance (~50Ω, calculated based on device geometry) as a result of the large tunnel junction capacitance, and hence generates only very small bias. Therefore we conclude that the observed RF-induced effects are bolometric.

Now we discuss the characterization of our devices as bolometers. We consider graphene as a radiation absorber which sits inside a heat bath of temperature $T_t$, being heated by 2 power sources: applied radiation $P_a$, and background radiation $P_b$. The electronic temperature of graphene is therefore: $T_e = T_t + f(P_a) + f(P_b)$, where $f(P)$ describes the heating of the electrons under the radiation power $P$. The junction resistance is a monotonic function of electronic temperature $R(T_e)$. In absence of the applied radiation ($P_a = 0$), the measured junction resistance is $R_1 = R(T_t+f(0)+f(P_b)) = R(T_t+f(P_b))$. On the other hand, at a base temperature $T_t=T_{t0}$, the measured junction resistance is $R_2 = R(T_{t0} + f(P_a)+f(P_b))$. By correlating these two cases so that $R_1=R_2$, we have: $T_t+f(P_b) = T_{t0}+ f(P_a)+f(P_b)$, hence $f(P_a) = T_t - T_{t0}$. We match the dynamic resistance equivalent of two cases: a). base temperature and under RF radiation; b). higher temperatures without radiation. In this way we are able to deduce the radiation induced electronic heating.

Following the model described above, we compare the two sets of measurements. We limit our discussion for the applied power between -90~-70dBm, where the radiation induced changes in *dV/dI* vs. $V_b$ curves match well with those at higher temperatures in absence of radiation. With this range, for each applied radiation power at the base temperature $T_{t0}$, we find the corresponding bath temperature $T_t$ in which the sample shows the same *dV/dI* vs. $V_b$ dependence. Figure 3a plots the relation between the applied RF power and the corresponding electronic heating $\Delta T = T_t - T_{t0}$, for two graphene superconducting tunnel junctions. From this measured relation, we can calculate the heat conductance of the graphene channel: $G=dP/dT$, shown in

Figure 3b. The thermal-noise-limited noise equivalent power (NEP) is $NEP = \sqrt{4k_B T^2 G}$ is estimated and plotted in Figure 3c.

Next we discuss the responsivity *dV/dP* of the devices. An order-of-magnitude estimation can be obtained by noticing that ~-60dBm radiation is required to smear out the superconducting tunnel junction features. Hence $dV/dP \sim \frac{\Delta/e}{-60dBm} \sim 10^5 V/W$. In more detailed calculations, we consider that the applied current is limited so that $V = IR \ll 2\Delta/e$. For our devices with $R \sim 5K\Omega$ at zero bias, we used an excitation current of $I = 10nA$, yielding $V \sim 50\mu V$ which is significantly small compared with $2\Delta/e \sim 200\mu V$. With the 10nA excitation data we calculate *dV/dP*, shown in Figure 3d. At low radiation power P < -70dBm ($10^{-10}$ W), the responsivity is ~$10^5$V/W. At higher radiation powers, the electron temperature in graphene rises close to the $T_c$ of aluminum, and the responsivity decreases rapidly.

Comparing our preliminary results with the state-of-the-art transition edge bolometer reported in ref [16, 17], the graphene-superconductor tunnel junction bolometers show significantly higher NEP as a result of large heat conductance. Possible electron cooling mechanisms include phonon cooling and diffusion cooling. At the range of temperatures and carrier densities studied in this work, electron-phonon scattering is well within the T<$T_{BG}$ regime where $T_{BG}$ is the Bloch-Gruneissen temperature. Based on reported measurements at high carrier density[18] of n>$10^{13}$ cm$^{-2}$ and theoretical expectations[19] that $\rho(T) = \alpha\, n^{-3/2}\, T^4$, we estimate electron-phonon scattering time at the conditions relevant to our measurements (T ~ 0.5K and n ~ $10^{12}$ cm$^{-2}$) that electron-phonon scattering time $\tau_{e\text{-}ph}$ ~ 30 μsec, leading to an estimated thermal conductance of $G_{e\text{-}ph} \sim C_e/\tau_{e\text{-}ph} \sim 10^{-17}$ W/K $\ll G_{measured}$. Given a typical mean free path of $l_{mfp} \sim 20 nm$, we estimate the electron-phonon scattering length of $l_{e-ph} = \sqrt{l_{mfp} v_F \tau_{e-ph}} \sim 800 \mu m$, much larger compared to sample size. We therefore conclude that cooling induced from electron-phonon scattering inside the graphene channel is insignificant in our devices. The main source of cooling is likely to be diffusion through the leads, because the tunnel barriers in our devices are non-ideal and "leaky". However preliminary estimation of heat conductance (based on Wiedemann-Franz law) yields values that are too small compared to the observations. Further study is needed in order to quantitative understand the observed large heat conductance, using devices with improved tunnel barrier and measurement setups which allow more accurate calibration of the RF power absorbed by the graphene channel. Another possible source of heat conductance is through substrate phonons. Transport measurements suggest that remote electron-phonon scattering from the substrate plays a significant role only for much higher temperatures (T>200K)[20]. However further detailed study of such cooling mechanism may be required, through comparison of bolometric response of graphene STJs on different substrates, and in absence of substrate (suspended graphene STJs). The response time of the graphene-aluminum bolometers measured here can be estimated from the measured heat conductance (~$2*10^{-10}$ W/K at ~200mK) and the calculated heat capacity (~ $10^{-21}$ J/K, shown in Figure 1b): $\tau = C/G \sim 5 ps$. Even at the cost of reducing the thermal conductance by 4 orders of magnitude, the response time is still within the $10^{-8}$ second regime. Therefore, with further improvements on tunnel barrier quality and possibly suspending the graphene absorbers, it is possible to achieve graphene STJ bolometers which are both ultrasensitive and ultrafast.

**Acknowledgement**

We gratefully thank Professor Daniel Prober and Dr. Daniel Santavicca for helpful discussions. We thank Professor Peter Stephens for providing the HOPG used in device fabrications. This work was supported by AFOSR-YIP award FA9550-10-1-0090.

**Figure Captions**

Figure 1. a. Circuit model for graphene STJ devices, based on which the power absorption in graphene can be estimated from the measured RF voltage/power on the 50Ω resistor. b. Calculated electron heat capacity and its temperature- and $V_g$ - dependence. The electron heat capacitance generally follows linear temperature dependence, except at the Dirac point where $C_e \sim T^2$ (however this is difficult to be observed experimentally since it only happens almost strictly at Dirac point). At low temperatures where thermal carrier excitation is negligible, the electron heat capacity depends linearly on Fermi energy, hence $\sqrt{V_g}$. At higher temperatures the thermal carrier excitation smears out such linear dependence.

Figure 2. a. Temperature dependence of the differential resistance vs. bias voltage curves, taken in absence of applied RF power. b. RF power (at 0.6GHz) dependence of the differential resistance vs. bias voltage curves taken at 160mK. The Inset shows the simulated RF response induced by non-linear IV characteristics, in absence of heating. c. Comparison between the bias dependence of differential resistance for 160mK/-70dBm and 500mK/no-RF. The two conditions yield almost identical results. d. Comparison between the bias dependence of differential resistance for 160mK/-60dBm and 925mK/no-RF. The broader peak feature measured at 160mK/-60dBm compared to that at 925mK/no-RF is a result of the reduced superconducting gap of the electrodes at higher bath temperature.

Figure 3. a. Temperature increase due to RF heating as a function of RF power. b. Temperature dependence of thermal conductance in the graphene STJ device. c. Temperature dependence of NEP. d. RF power dependence of responsivity.

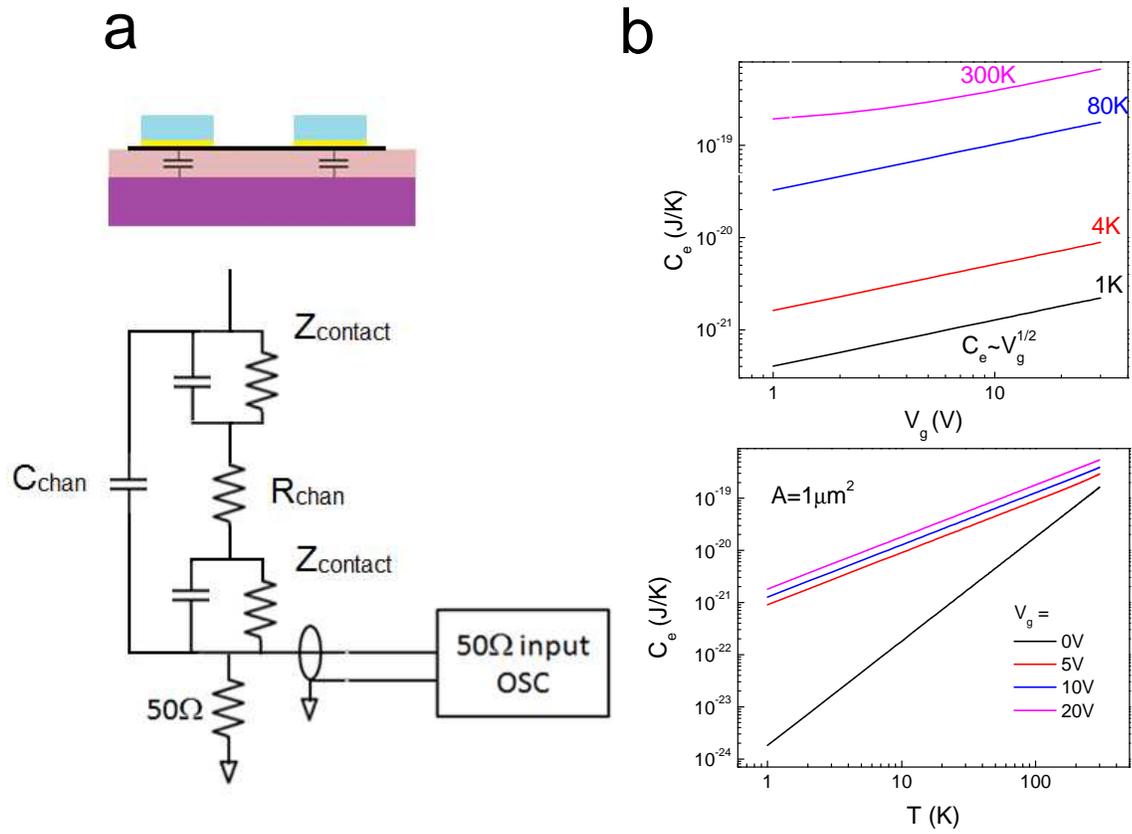

Figure 1.

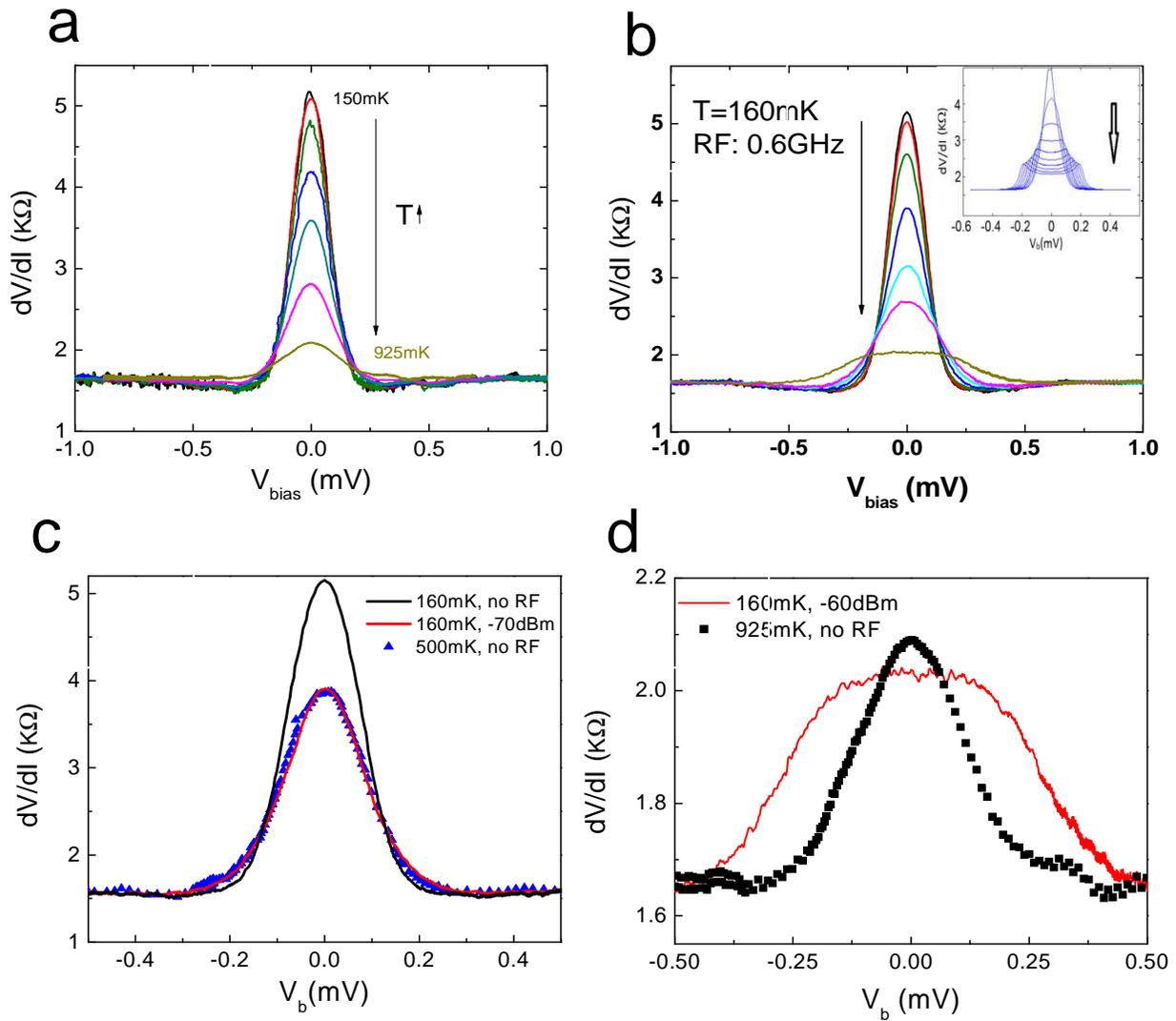

Figure 2.

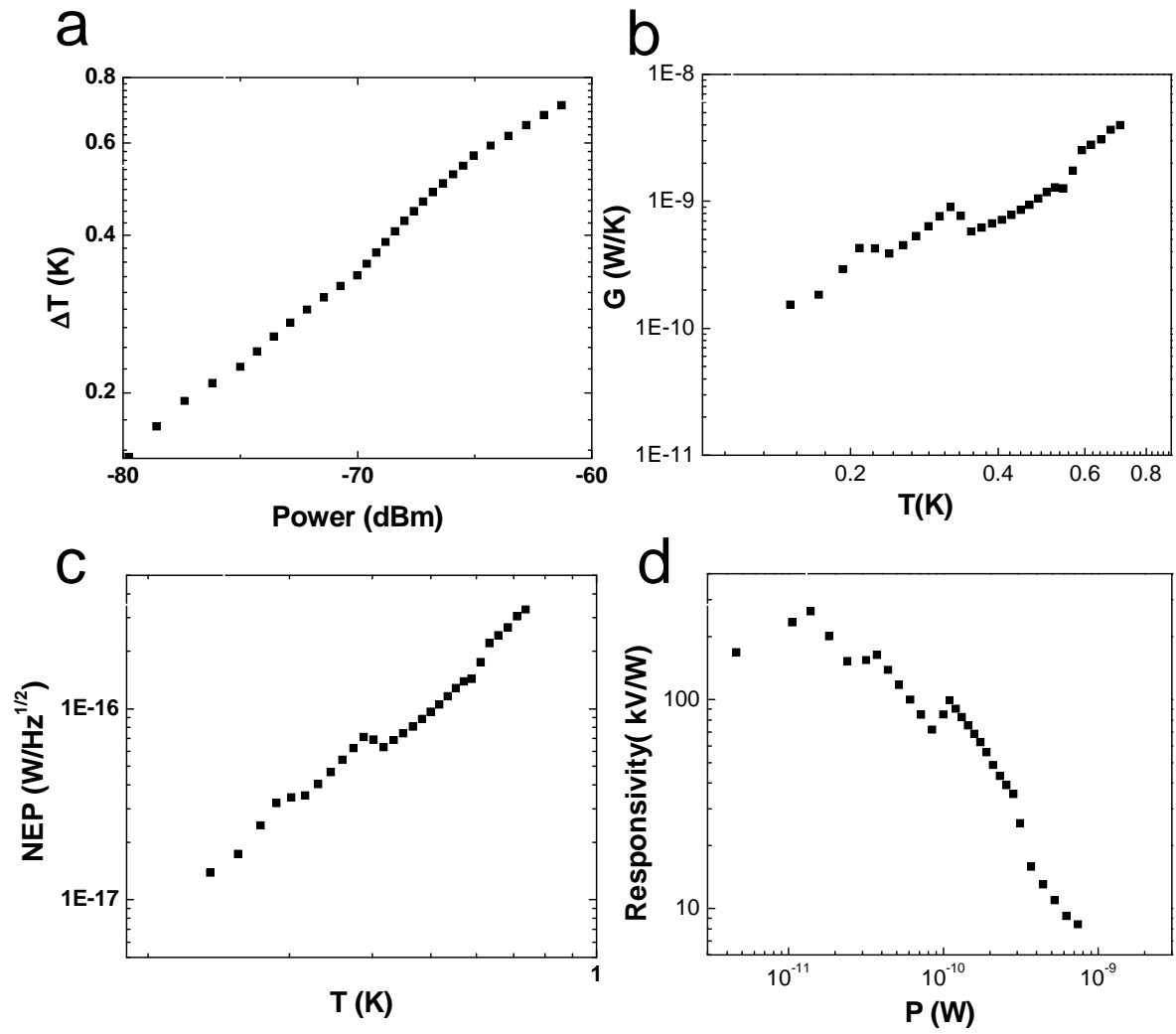

Figure 3.